\documentclass[11pt]{article}
\usepackage[latin1]{inputenc}
\usepackage{colortbl}
\usepackage{color}
\usepackage{array}
\usepackage{amsmath, amssymb, bm, epsfig}
\usepackage[a4paper, text={17cm,22.5cm},top=2cm, left=3.5cm, right=2.5cm, bottom=2cm, includeheadfoot, headheight=0.5cm]{geometry}
\usepackage{longtable}
\usepackage{tabularx,ragged2e,booktabs,caption}
\usepackage{graphicx,epstopdf}

\begin{document}
\title{\textbf{A reduced new modified Weibull distribution}}
\date{}
\author{Saad J. Almalki\\
School of Mathematics, University of Manchester, Manchester M13 9PL, UK}
\maketitle

\begin{abstract}
In this paper, we propose a reduced version of the new modified Weibull (NMW) distribution
due to Almalki and Yuan \cite{meNMW} in order to avoid some estimation problems.
The number of parameters in the NMW distribution is five.
The number of parameters in the reduced version is three.
We study mathematical properties as well as  maximum likelihood estimation of the reduced version.
Four real data sets (two of them complete and the other two censored)
are used to compare the flexibility of the reduced version versus the NMW distribution.
It is shown that the reduced version has the same desirable properties of the NMW distribution
in spite of having two less parameters.
The NMW distribution did not provide a significantly better fit than the reduced version
for any of the four data sets.
\end{abstract}


\section{Introduction}

In reliability engineering and lifetime analysis many applications require a bathtub shaped hazard rate function.
The traditional Weibull distribution \cite{Weibull}, which includes the exponential and Rayleigh distributions as particular cases,
is one of the important lifetime distributions.
Unfortunately, however, it does not exhibit a bathtub shape for its hazard rate function.
Many researchers have proposed modifications and generalizations of the
Weibull distribution to accommodate bathtub shaped hazard rates.
Extensive reviews of these modifications have been presented by many authors, see, for example,
Rajarshi and Rajarshi \cite{Rajarshi}, Murthy {\em et al.} \cite{Murthy},
Pham and Lai \cite{Pham_and_Lai} and Lai {\em et al.} \cite{LaiWeibulldists}.

Although some flexible distributions among these modifications have only two or three parameters
(for example, the flexible Weibull extension \cite{Bebbingtonet-FlexW},
the modified Weibull (MW) distribution \cite{Lai_MW}
and the modified Weibull extension \cite{XieMWextension}),
most modifications of the Weibull  distribution have four or five parameters.
By combining two Weibull distributions, with one having an increasing hazard rate function
and the other decreasing one, Xie and Lai \cite{Xie_Lai_addW} introduced a four-parameter
distribution called the additive Weibull (AddW) distribution.
Sarhan and Apaloo \cite{MWEx_Sarhan13} proposed the exponentiated modified Weibull extension
which exhibits a bathtub shaped hazard rate.
Sarhan {\em et al.} \cite{Sarh13_ExGLenE} proposed the exponentiated generalized linear exponential
distribution which generalizes a large set of distributions including the exponentiated Weibull distribution.
Famoye {\em et al.} \cite{20-BWD} proposed the beta-Weibull (BWD) distribution with unimodal,
increasing, decreasing or bathtub shaped hazard rate functions.
Another four-parameter distribution, called the generalized modified Weibull (GMW) distribution,
was proposed by Carrasco {\em et al.} \cite{Carrasco_GMW}.
The hazard rate function of this distribution can be increasing, decreasing, bathtub shaped or unimodal.
Cordeiro {\em et al.} \cite{24-KumW} introduced the Kumaraswamy Weibull (KumW) distribution and
studied its mathematical properties extensively.
It has four parameters, three of which are shape parameters, making it so flexible.
Its hazard function can be constant, increasing, decreasing, bathtub shaped or unimodal.
A five-parameter distribution was introduced as a modification of the Weibull distribution by Phani \cite{PhaniMW}.
This distribution generalizes the four-parameter Weibull distribution proposed by Kies \cite {KiesMW58}.
Another five-parameter distribution, the beta modified Weibull (BMW) distribution,
was introduced by Silva {\em et al.} \cite{Silva_BMW}.
It allows for different hazard rate shapes: increasing, decreasing, bathtub shaped and unimodal.
It was shown to fit bathtub shaped data sets very well.
Recently, a new five-parameter distribution called the beta generalized Weibull distribution was
proposed by Singla {\em et al.} \cite{BGW}.
The hazard rate function of this distribution can be increasing, decreasing, bathtub shaped or unimodal.
It contains as sub-models some well known lifetime distributions.

Almalki and Yuan \cite{meNMW} introduced a new modified Weibull (NMW) distribution.
It generalizes several commonly used distributions in reliability and lifetime data analysis,
including the MW distribution, the AddW distribution,
the modified Weibull distribution of Sarhan and Zaindin (SZMW) \cite{SZ_MW}, the Weibull distribution,
the exponential distribution, the Rayleigh distribution,
the extreme-value distribution and the linear failure rate (LFR) \cite{LFRD} distribution.
The cumulative distribution function (CDF) of the NMW distribution is
\begin{eqnarray*}
F(x)=1-e^{-\alpha x^{\theta} -\beta x^{\gamma}e^{\lambda x}}
\end{eqnarray*}
for $x > 0$, $\gamma > 0$, $\theta > 0$, $\alpha > 0$, $\beta > 0$ and $\lambda>0$,
where $\gamma$, $\theta$ are shape parameters,
$\alpha$, $\beta$ are scale parameters and $\lambda$ is an acceleration parameter.
Almalki and Yuan \cite{meNMW} derived mathematical properties of this distribution as well as
estimated its parameters by the method of maximum likelihood with application to real data sets.

The hazard rate function of the NMW distribution can be increasing, decreasing or bathtub shaped.
It has been shown to be the best lifetime distribution to date in
terms of fitting some popular and widely used real data sets like Aarset data \cite{Aarset} and voltage data \cite{Meeker}.

Although distributions with four or more parameters are flexible and
exhibit bathtub shaped hazard rates, they are also
complex \cite{Nelson} and cause estimation problems as a consequence
of the number of parameters, especially when the sample size is not large.
The main purpose of this work is to reduce the number of parameters of
the NMW distribution so as to address these problems while maintaining the same
flexibility to fit data so well.
This can be achieved by choosing the two shape parameters $\gamma=\theta=\frac{1}{2}$ as we now show.


There are various tools to assess the flexibility of a given univariate distribution.
One commonly used tool is the kurtosis-skewness plot.
Values of (skewness, kurtosis) plotted on the $(x, y)$ plane
for all possible values of the parameters of the distribution
give what is referred to as the kurtosis-skewness plot.
The area or the range covered by the kurtosis-skewness plot is a measure of flexibility of the distribution.

The kurtosis-skewness plot for the NMW distribution is drawn on the left hand side of Figure .
The values of (skewness, kurtosis) were computed over
$\theta = 0.1, 0.2, \ldots, 5$, $\gamma = 0.1, 0.2, \ldots, 5$,
$\theta = \gamma \neq \frac{1}{2}$,
$\alpha = 0.1, 0.2, \ldots, 2$, $\beta = 0.1, 0.2, \ldots, 2$ and $\lambda = 0.1, 0.2, \ldots, 2$.
The kurtosis-skewness plot for the reduced distribution is drawn on the right hand side of Figure .
The values of (skewness, kurtosis) were computed over
$\alpha = 0.1, 0.2, \ldots, 2$, $\beta = 0.1, 0.2, \ldots, 2$ and $\lambda = 0.1, 0.2, \ldots, 2$.

We can see that the particular case $\theta = \gamma = \frac{1}{2}$ is a flexible member of the NMW distribution.
The range of kurtosis values is widest for the case $\theta = \gamma = \frac{1}{2}$.
The range of skewness values is also widest for the case $\theta = \gamma = \frac{1}{2}$,
but some negative skewness values are not accommodated by this case.

The rest of this paper is organized as follows.
The reduced distribution is introduced in Section 2.
Section 3 considers the hazard rate function of the reduced distribution.
The moments,  the moment generating function and the distribution of order statistics
are derived as particular cases of the NMW distribution in Section 4.
Section 5 discusses maximum likelihood estimation of the unknown parameters using
complete and censoring data.
Four real data sets, uncensored and censored, are analyzed in Section 6.
Finally, Section 7 concludes the paper.

\section{The reduced distribution}

Setting $\gamma=\theta=\frac{1}{2}$ in the CDF of the NMW distribution,
we obtain the CDF of the reduced new modified Weibull (RNMW) distribution  as
\begin{eqnarray*}
F(x)=1-e^{-\alpha \sqrt{x}-\beta \sqrt{x}e^{\lambda x}}
\end{eqnarray*}
for $x > 0$, $\alpha > 0$, $\beta > 0$ and $\lambda>0$,
where $\alpha$, $\beta$ are scale parameters and $\lambda$ is an acceleration parameter.
This reduced version of the NMW distribution has a bathtub shaped hazard rate function, as will be shown later.

\begin{figure}[here]
 \centering
    \includegraphics[width=11cm,height=7.5cm]{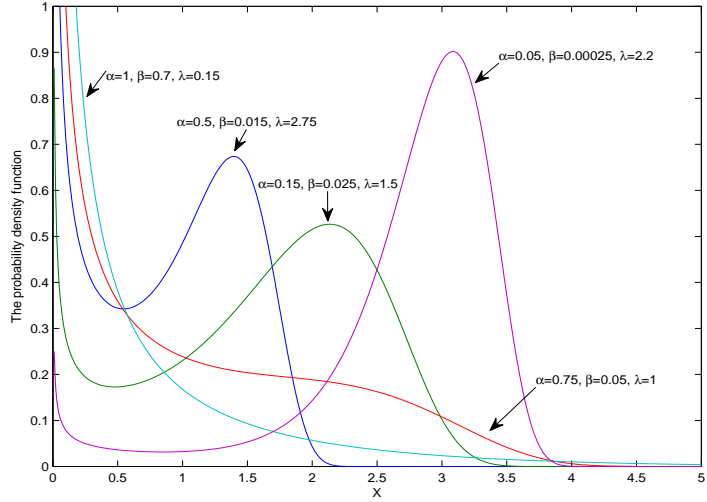}\\
  \caption{PDFs of the RNMW distribution.}
\label{pdfsNMW}
\end{figure}

The corresponding probability density function (PDF) is
\begin{eqnarray*}
f(x)=\frac{1}{2\sqrt{x}}\left[ \alpha +\beta (1 +2\lambda x)e^{\lambda x}\right] e^{-\alpha\sqrt{x}-\beta \sqrt{x}e^{\lambda x}}
\end{eqnarray*}
for $x > 0$.
The PDF of the RNMW distribution can be decreasing or decreasing then increasing-decreasing
as shown in Figure \ref{pdfsNMW}.
It is clear that the reduced distribution is nearly as flexible as the NMW distribution.

\section{The hazard rate function}

\begin{figure}[here]
 \centering
\begin{tabular}{cc}
  \includegraphics[width=7.5cm]{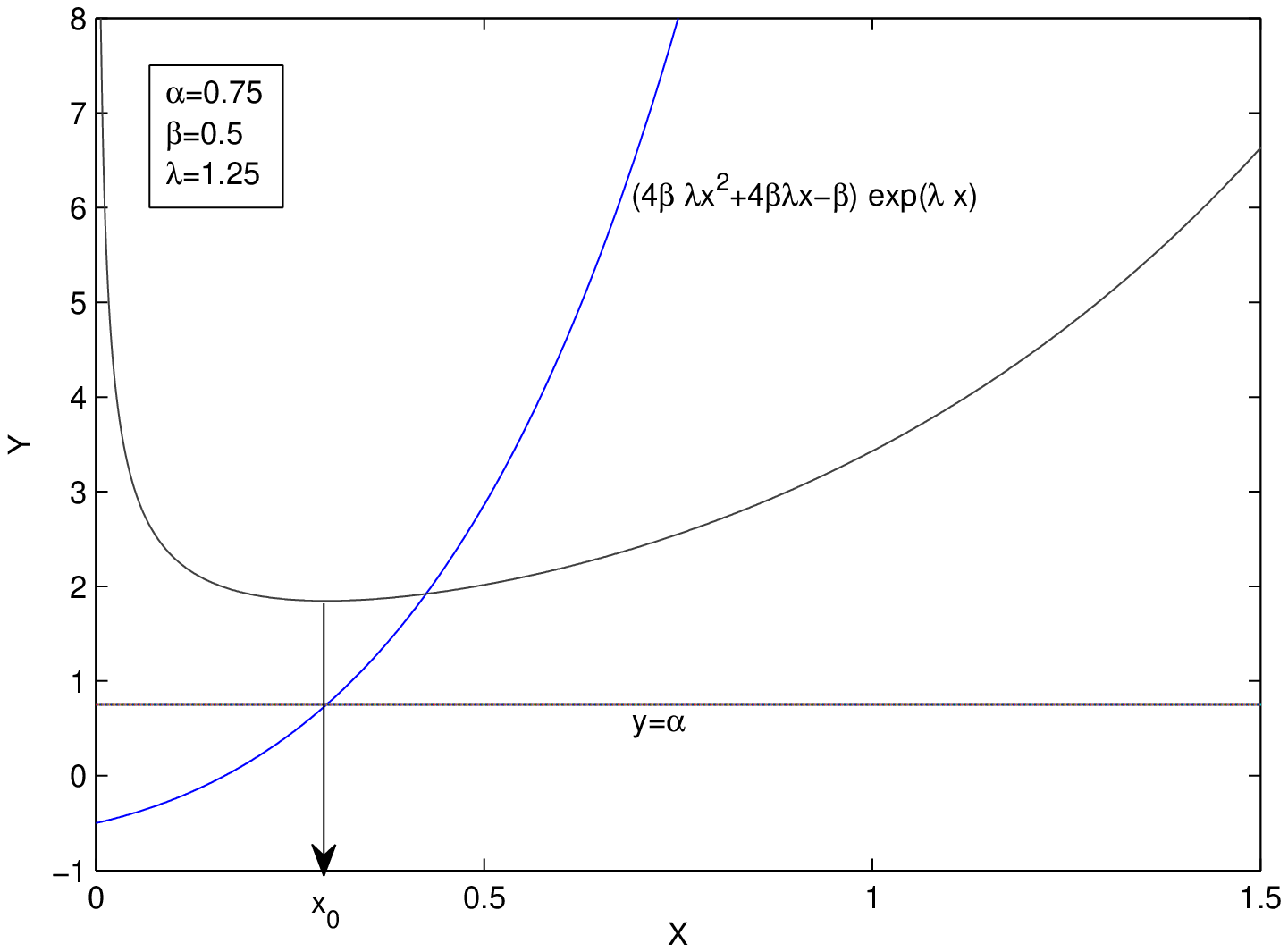} &
  \includegraphics[width=7.5cm]{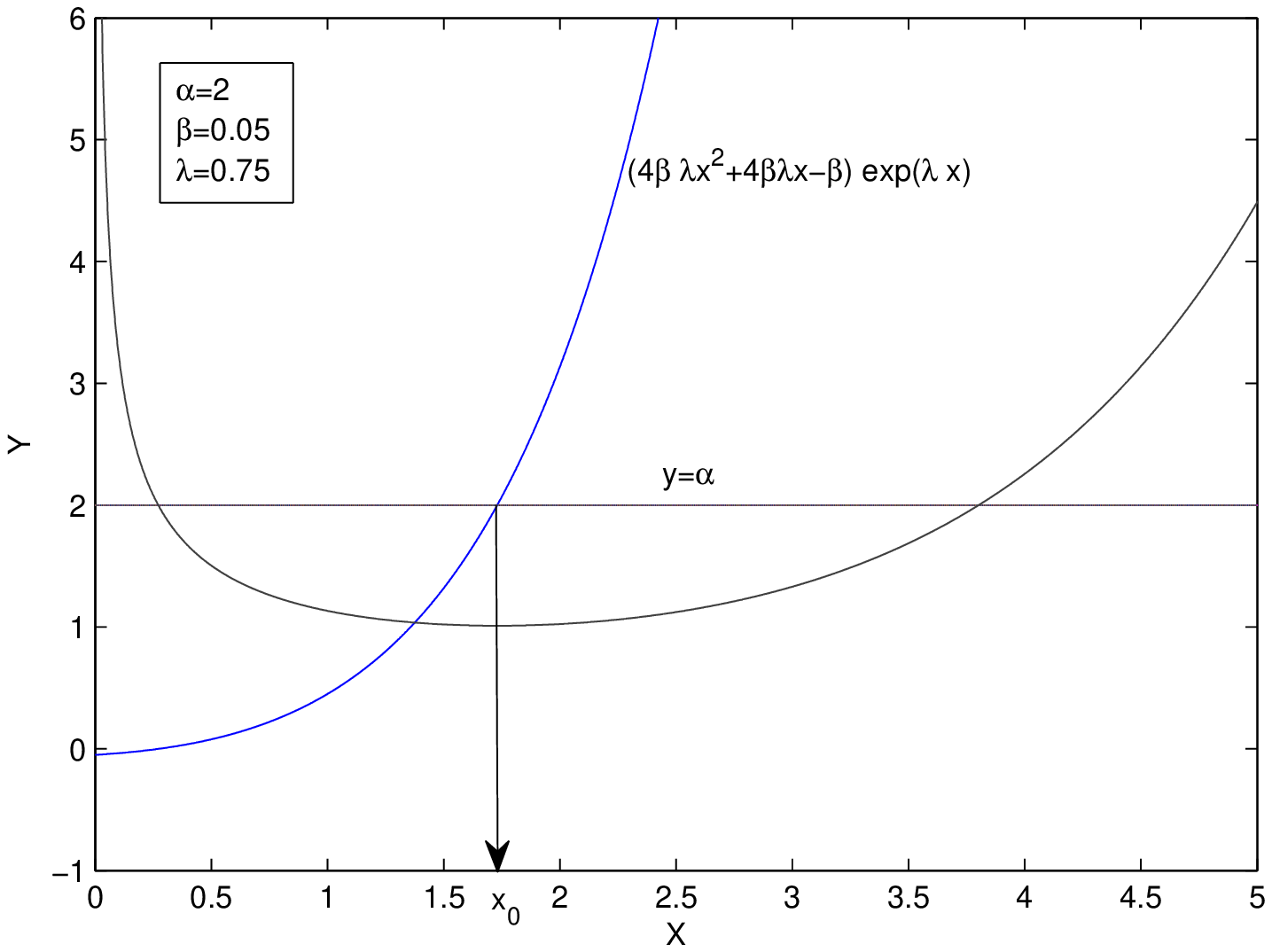}\\
  (a) & (b)\\
  \end{tabular}
\caption{The root of $\frac{d}{dx}\log\left\{{h(x_{0})}\right\}=0$.}
\label{root_dh}
\end{figure}

The hazard rate function of the RNMW distribution is
\begin{eqnarray}
h(x) = \frac{1}{2\sqrt{x}}\left[ \alpha +\beta (1 +2\lambda x)e^{\lambda x}\right]
\label{h}
\end{eqnarray}
for $x > 0$.
To derive the shape of $h(x)$, we obtain the first derivative of $\log\left\{{h(x)}\right\}$:
\begin{eqnarray}
\frac{d}{dx}\log\left\{{h(x)}\right\}=-\frac{1}{2x}+\frac{\beta \lambda(2\lambda x+3)e^{\lambda x}}
{\alpha+\beta(2\lambda x+1)e^{\lambda x}}.
\label{d_h}
\end{eqnarray}
Setting this to zero, we have
\begin{eqnarray}
-\frac{1}{2x}+\frac{\beta \lambda(2\lambda x+3)e^{\lambda x}}{\alpha+\beta(2\lambda x+1)e^{\lambda x}} = 0.
\label{d_h_0}
\end{eqnarray}
Let $x_{0}$ denote the root of (\ref{d_h_0}).
From (\ref{d_h}), $\frac{d}{dx}\log\left\{{h(x)}\right\}<0$ for $x\in(0,x_{0})$,
$\frac{d}{dx}\log\left\{{h(x_{0})}\right\}=0$, and $\frac{d}{dx}\log\left\{{h(x)}\right\}>0$ for $x>x_{0}$.
So, $h(x)$  initially decreases before increasing.
Hence, we have a bathtub shape.
Let $x_{0}$ denote the solution of (\ref{d_h_0}); that is, the solution of
\begin{eqnarray}
\left(4\beta \lambda x^2+4\beta \lambda x-\beta\right)e^{\lambda x}=\alpha.
\label{d_h_0_2}
\end{eqnarray}
The left hand side of (\ref{d_h_0_2}) is  $e^{\lambda x}$ multiplied by the
quadratic function $(4\beta\lambda x^2+4\beta\lambda x-\beta)$.
The value $x_0$ is unique and positive as shown in Figure \ref{root_dh}.

\begin{figure}[here]
 \centering
 \begin{tabular}{cc}
  \includegraphics[width=7.85cm]{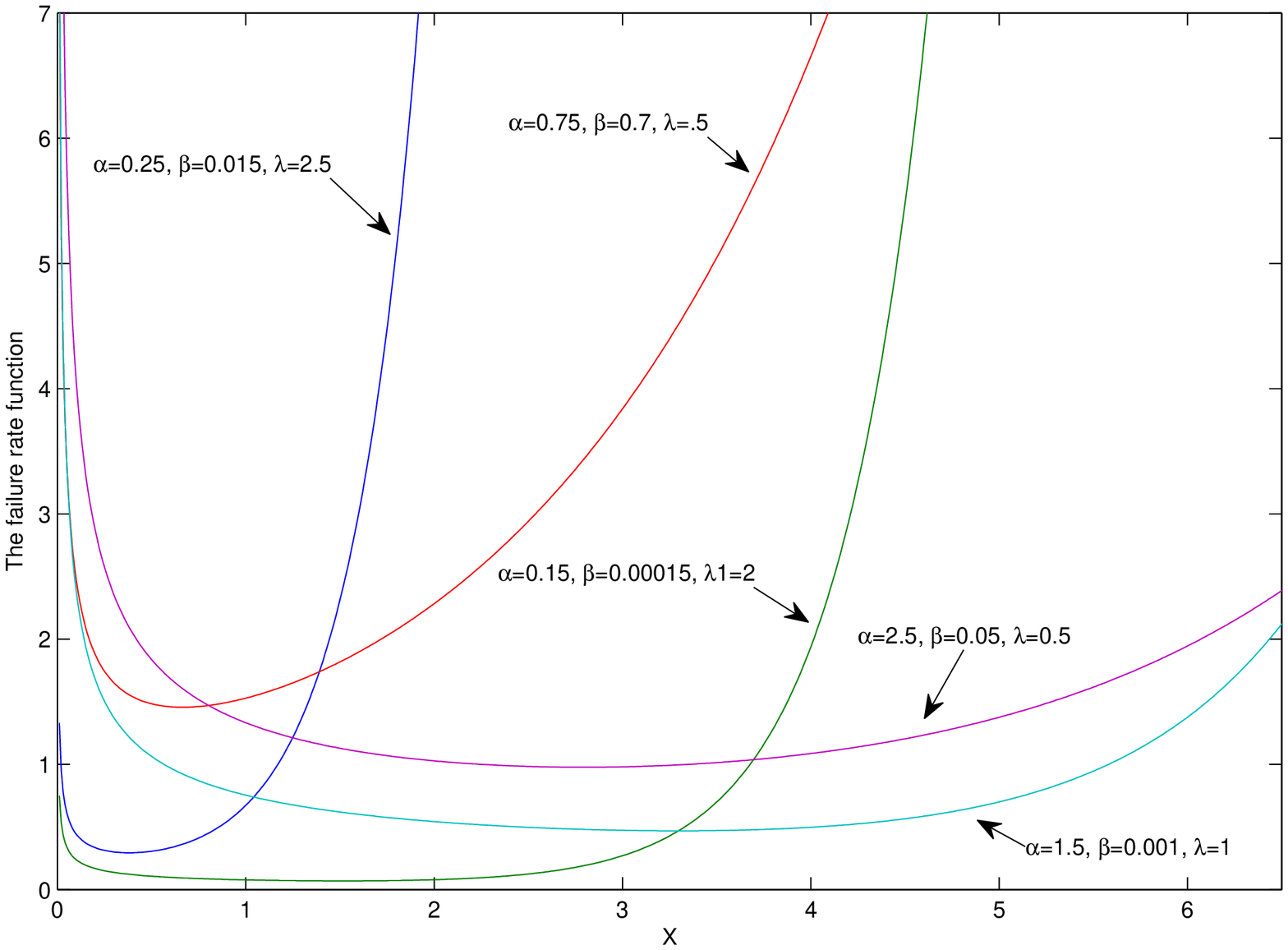} &
  \includegraphics[width=7.15cm]{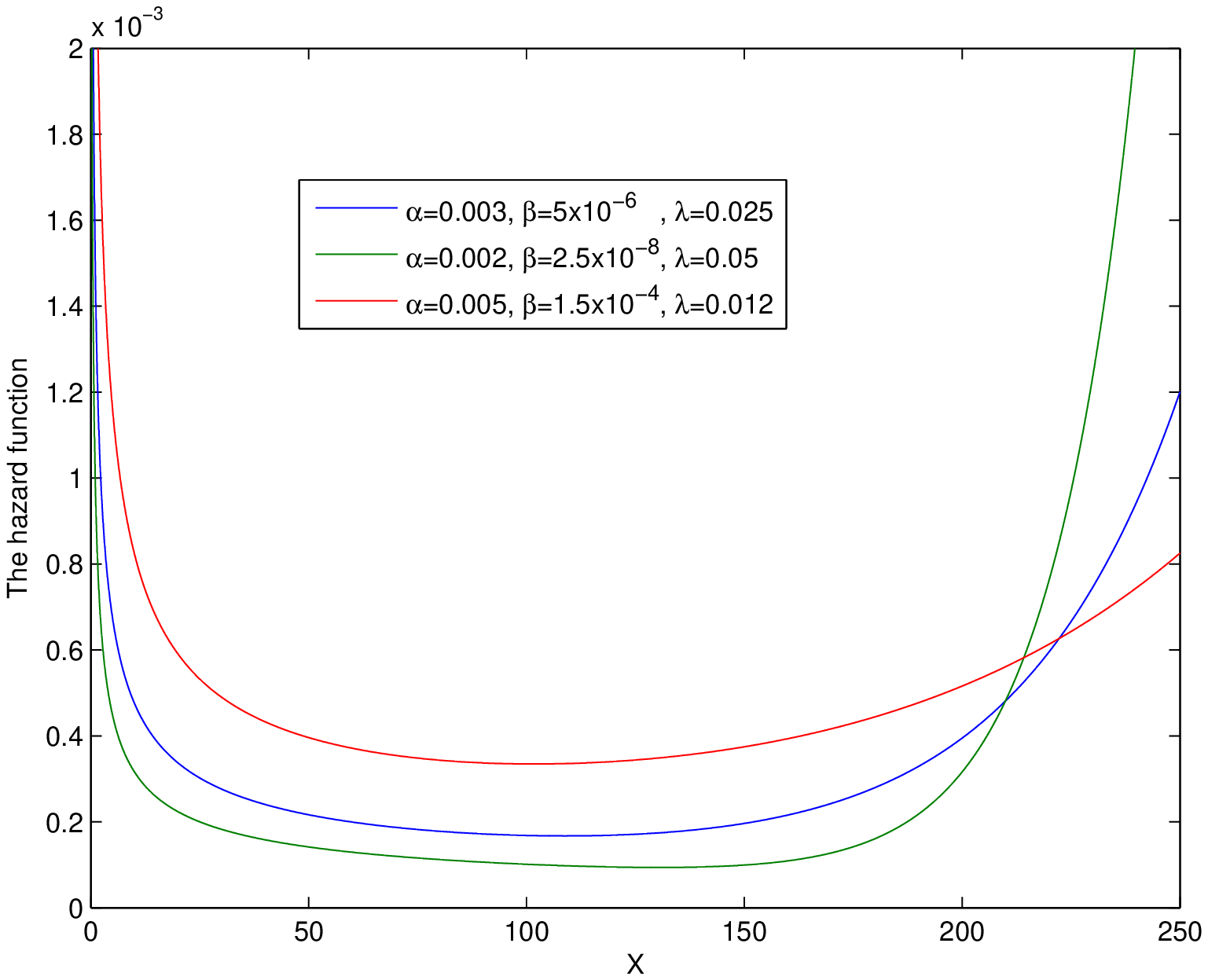}\\
   (a) & (b)\\
   \end{tabular}
  \caption{Hazard rate function of the RNMW distribution.}
\label{hazrdsNMW}
\end{figure}

Plots of the hazard rate function of the RNMW distribution are shown in Figure \ref{hazrdsNMW} (a).
Many different applications in reliability and lifetime analysis require
bathtub shaped hazard rate functions with a long useful life period and the middle period
of the bathtub shape having a relatively constant hazard rate (for example, electric
machine life cycles and electronic devices \cite{optimal_Rel}).
The NMW distribution has this property, and so does the RNMW distribution, as Figure \ref{hazrdsNMW}(b) shows.

\section{The moments, the moment generating function and order statistics}

Let $X$ denote a random variable having the RNMW distribution.
The $r$th moment of $X$ can
be derived from Section 3.2 in  Almalki and Yuan \cite{meNMW} as
\begin{eqnarray}
\mu_{r}^{\prime}=2r\sum_{n,m=0}^{\infty}
\frac {(-\beta)^{n}(n\lambda)^{m}}
{n!m!}\frac {\Gamma \left( n+2(m+r)\right)}{\alpha^{n+2(m+r)}}
\label{moment}
\end{eqnarray}
for $r=1,2,\ldots$.
The moment generating function of $X$ is
\begin{eqnarray}
M_{X}(t)=1+2\sum_{n,m,k=0}^{\infty}\frac {(-\beta)^{n}(n\lambda)^{m}t^{k+1}}{i!j!k!}
\left[ \frac {\Gamma \left( n+2(m+k)+2\right)}{\alpha^{n+2(m+k)+2}} \right],
\label{MGF_RNM2}
\end{eqnarray}
see Appendix A for a proof.
Using (\ref{MGF_RNM2}), the first four moments of $X$ are
\begin{eqnarray*}
&&
M_{X}^{\prime}(0) = \mu_{1}^{\prime}=2\sum_{n,m=0}^{\infty}
\frac{(-\beta)^{n}(n\lambda)^{m}}{i!j!k!}\left[ \frac{\Gamma \left(
n+2m+2\right)}{\alpha^{n+2m+2}}\right],
\\
&&
M_{X}^{\prime \prime}(0) = \mu_{2}^{\prime \prime}=4\sum_{n,m=0}^{\infty}
\frac{(-\beta)^{n}(n\lambda)^{m}}{i!j!k!}\left[ \frac{\Gamma \left(
n+2m+4\right)}{\alpha^{n+2m+4}}\right],
\\
&&
M_{X}^{(3)}(0) = \mu_{3}^{(3)}=6\sum_{n,m=0}^{\infty}
\frac{(-\beta)^{n}(n\lambda)^{m}}{i!j!k!}\left[ \frac{\Gamma \left( n+2m+6\right)}{\alpha^{n+2m+6}}\right],
\\
&&
M_{X}^{(4)}(0) = \mu_{2}^{(4)}=8\sum_{n,m=0}^{\infty}
\frac{(-\beta)^{n}(n\lambda)^{m}}{i!j!k!}\left[ \frac{\Gamma \left( n+2m+8\right)}{\alpha^{n+2m+8}}\right].
\end{eqnarray*}
These expressions are consistent with the formula for moments in (\ref{moment}).

Let $X_1, \ldots, X_n$ denote a random sample
drawn from the RNMW distribution
with parameters $\alpha$, $\beta$ and $\lambda$.
The PDF of the $r$th order statistic say $X_{(r)}$
can be derived as a particular case from Section 3.3 of \cite{meNMW}:
\begin{eqnarray*}
{f}_{r:n}\left( x\right)=n\binom{n-1}{r-1}
\sum\limits_{{\ell=0}}^{{r-1}}\binom{r-1}{\ell}
\frac {(-1)^{\ell}}{(n+\ell+1-r)}f \left( x; \alpha_\ell, \beta_\ell, \lambda \right),
\end{eqnarray*}
where $f(x; \alpha_\ell, \beta_\ell, \lambda)$ is the RNMW PDF with parameters $\alpha_{\ell}=(n+\ell+1-r)\alpha $,
$\beta_{\ell}=(n+\ell+1-r)\beta$ and $\lambda$.
The $k$th non-central moment of the $r$th order statistic is then
\begin{eqnarray*}
{\mu}_k^{'(r:n)}={2nk}\binom{n-1}{r-1}\sum_{i,j=0}^{\infty}
\sum\limits_{{\ell=0}}^{{r-1}}
\binom{r-1}{\ell}\frac {(-1)^{\ell+1}\beta^{i}(i\lambda)^{j}\Gamma \left( i +2(j+k)\right)}
{\alpha^{i-1} \left[ \alpha (n+\ell+1-r) \right]^{{2(j+k)}+1}}.
\end{eqnarray*}

\section{Parameter estimation}

In this section, point and interval estimators of the unknown parameters of
the RNMW distribution are derived using the maximum likelihood method.
We consider both complete data and censored data.

\subsection{Complete data}

The PDF of the RNMW distribution can be rewritten as
\begin{eqnarray*}
f(x)=h \left(x;\underline{\vartheta}\right) e^{-\alpha \sqrt{x}-\beta \sqrt{x}e^{\lambda x}}
\end{eqnarray*}
for $x > 0$, where $h\left( x; \underline{\vartheta} \right)$ is the hazard rate function
in (\ref{h}) and $\underline{\vartheta}=(\alpha, \beta, \lambda)$ is a vector of parameters.

Let $x_1, \ldots, x_n$ denote a random sample of complete data from the RNMW distribution.
Then, the log-likelihood function is
\begin{eqnarray*}
\ell \left( \underline{\vartheta} \right) = \sum\limits_{i=1}^{n} \left[ \log \left(h \left( x_{i};
\underline{\vartheta} \right) \right)-\alpha \sqrt{x_{i}}-\beta \sqrt{x_{i}}e^{\lambda x_{i}}\right].
\end{eqnarray*}
The likelihood equations are
obtained by setting the first partial
derivatives of $\ell$ with respect to $\alpha, \beta$ and $\lambda $ to zero; that is,
\begin{eqnarray}
&&
\sum\limits_{i=1}^{n}\frac{1}{h \left( x_i; \alpha, \beta,\lambda \right)
\left( 2\sqrt{x_{i}} \right)} - \sum\limits_{i=1}^{n}{\sqrt{x_{i}}} = 0,
\label{a=o}
\\
&&
\sum\limits_{i=1}^{n}
\frac {\left( 0.5 +\lambda x_{i} \right) e^{\lambda x_{i}}}
{h \left( x_i; \alpha, \beta, \lambda \right)\sqrt{x_{i}}} -
\sum\limits_{i=1}^{n}e^{\lambda x_{i}}\sqrt{x_{i}} = 0,
\label{b=0}
\\
&&
\sum\limits_{i=1}^{n}\frac {\sqrt{x_{i}} \left( \frac{3}{2} +\lambda x_{i} \right) e^{\lambda x_{i}}}
{h \left( x_{i}; \alpha, \beta, \lambda \right)} -
\sum\limits_{i=1}^{n}e^{\lambda x_{i}}\sqrt{x_{i}^{3}} =0.
\label{lambda=0}
\end{eqnarray}

\subsection{Censored data}

Here, we consider maximum likelihood estimation for censored data without replacement.
Let  $X_{i}$ and $C_{i}$ denote the lifetime and the censoring time for tested individual $i$, $i = 1, \ldots, n$.
Suppose $X_{i}$ and $C_{i}$ are independent random variables.
The failure times are $x_{i} = \min(X_{i},C_{i})$, $i = 1, \ldots, n$.
Then, the log-likelihood function is
\begin{eqnarray*}
\ell \left( \underline{\vartheta} \right) = \sum\limits_{i=1}^{d} \log \left[ h \left(x_{i};
\underline{\vartheta} \right) - \alpha \sqrt{x_{i}}-\beta \sqrt{x_{i}}e^{\lambda x_{i}}\right] -
\sum\limits_{i\in C}\left[ \alpha \sqrt{x_{i}}+\beta \sqrt{x_{i}}e^{\lambda x_{i}}\right],
\end{eqnarray*}
where $d$ is the number of failures and $C$ indexes the censored observations.

Setting the first partial derivatives of $\ell \left( \underline{\vartheta} \right)$ with
respect to $\alpha$, $\beta$ and $\lambda $ to zero, the likelihood equations are obtained as
\begin{eqnarray}
&&
\sum\limits_{i=1}^{d}\left\{
\frac {1}{h \left( x_i; \alpha, \beta, \lambda \right) \left( 2\sqrt{x_{i}} \right)} - \sqrt{x_{i}} \right\} -
\sum\limits_{i\in C}\sqrt{x_{i}} = 0,
\label{a2=o}
\\
&&
\sum\limits_{i=1}^{d}\left\{
\frac {\left( 0.5 +\lambda x_{i} \right) e^{\lambda x_{i}}}{h \left( x_i; \alpha, \beta, \lambda \right) \sqrt{x_{i}}} -
e^{\lambda x_{i}}\sqrt{x_{i}} \right\} -
\sum\limits_{i\in C}\sqrt{x_{i}}e^{\lambda x_{i}} = 0,
\label{b2=0}
\\
&&
\sum\limits_{i=1}^{d}\left\{
\frac {\sqrt{x_{i}} \left( \frac{3}{2} + \lambda x_{i} \right)
e^{\lambda x_{i}}}{h \left( x_{i}; \alpha, \beta, \lambda \right)} -
e^{\lambda x_{i}}\sqrt{x_{i}^{3}} \right\} - \sum\limits_{i\in C}e^{\lambda x_{i}}\sqrt{x_{i}^{3}} = 0.
\label{lambda2=0}
\end{eqnarray}

By solving the  systems of nonlinear likelihood equations,
(\ref{a=o},\ref{b=0},\ref{lambda=0}) and (\ref{a2=o},\ref{b2=0},\ref{lambda2=0}),
numerically for $\alpha$, $\beta$ and $\lambda$,
we can obtain maximum likelihood estimates for complete and censored data.

According to Miller \cite{Miller}, the MLEs ($\widehat{\alpha}, \widehat{\beta}, \widehat{\lambda}$)
of ($\alpha, \beta, \lambda$) have an approximate multivariate normal
distribution with mean ($\alpha$, $\beta$, $\lambda $) and
variance-covariance matrix $\widehat{I^{-1}}$; that is,
\begin{eqnarray*}
\left( \widehat{\alpha}, \widehat{\beta}, \widehat{\lambda} \right)
\sim
N_{3} \left( (\alpha, \beta, \lambda), \widehat{I^{-1}} \right),
\end{eqnarray*}
where
\begin{eqnarray*}
I=-\left[
\begin{array}{lll}
\ell_{\alpha \alpha} & \ell_{\alpha \beta} & \ell_{\alpha\lambda}
\\
& \ell_{\beta \beta}& \ell_{\beta \lambda}
\\
&  & \ell_{\lambda \lambda}
\end{array}
\right].
\end{eqnarray*}
The second order partial derivatives of $\ell \left(\underline{\vartheta}\right)$ are given in Appendices B and C.

\section{Applications}

This section provides four applications,
two of them are for complete (uncensored) data sets
and the others are for censored data sets,
to show how the RNMW distribution can be applied in practice.
Almalki and Yuan \cite{meNMW} have shown that the NMW distribution fits data sets better than
existing modifications of the Weibull distribution like the BMW distribution,
the AddW distribution, the MW distribution and the SZMW distribution.
So, we shall compare the fits of the RNMW and NMW distributions
to see if the former can perform as well as the NMW distribution.
The Kolmogorov-Smirnov (K-S) statistic (the distance
between the empirical CDFs and the fitted CDFs), the Akaike information criterion (AIC),
the Bayesian information criterion (BIC) and the consistent Akaike information criterion (CAIC)
are used to compare the candidate distributions.
The log-likelihood ratio test is used to compare the NMW and RNMW distributions by
testing the hypotheses: $H_{0}:\theta=\gamma=\frac{1}{2}$ versus $H_{1}: H_{0}$ is false.
The likelihood ratio test statistic
for testing $H_{0} $ against $H_{1}$ is $\omega=2(\mathcal{L}_{NMW}-\mathcal{L}_{RNMW})$,
which follows a $\chi^{2}$ distribution with two degrees of freedom under $H_0$.

\subsection{Complete data}

In this section, we show how the RNMW distribution can be applied in practice for two complete (uncensored) real data sets.

\subsubsection{Aarset data}

\begin{figure}
 \centering
\begin{tabular}{cc}
\includegraphics[width=7.25cm]{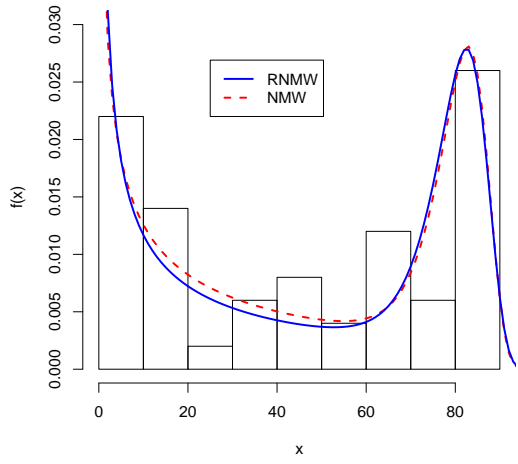} &
  \includegraphics[width=7cm]{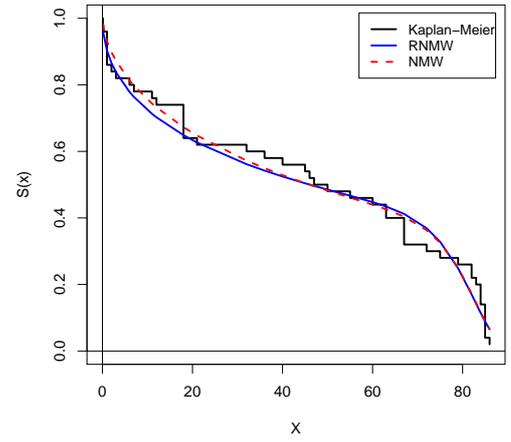}\\
  (a) & (b)\\
  \includegraphics[width=8.5cm]{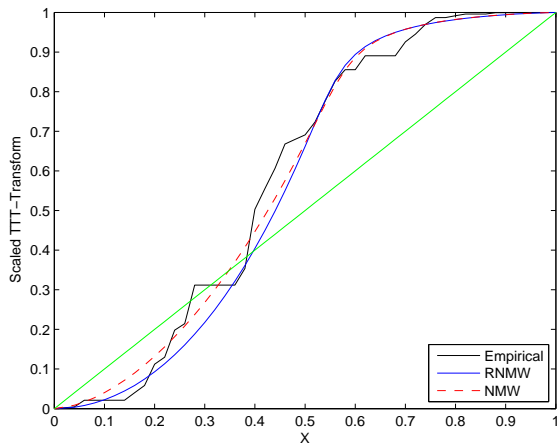} &
  \includegraphics[width=7cm]{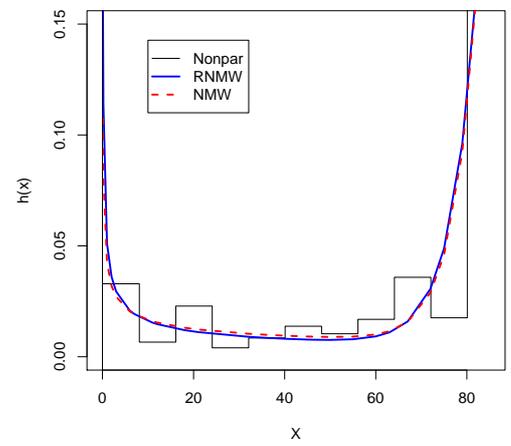}
  \\
  (c) & (d)
  \end{tabular}
\caption{For Aarst data: (a) Histogram and fitted PDFs;
(b) Empirical and fitted survival functions;
(c) Empirical and fitted TTT-transforms;
(c) Nonparametric and fitted hazard rate functions.}
\label{a1234}
\end{figure}

The Aarset data \cite{Aarset} consisting of lifetimes of fifty devices
is widely used in lifetime analysis.
The data set exhibits a bathtub shaped hazard rate.
Both the NMW and RNMW distributions were fitted to this data set.
Table \ref{MLEs Aarst} gives
the MLEs of the parameters, corresponding standard errors, AIC, BIC,
and CAIC.
Table \ref{KS_AIC_Aarst} provides the K-S test statistics.
Figures \ref{a1234}a and \ref{a1234}b show the histogram of the data, PDFs of the fitted NMW and RNMW distributions,
the empirical survival function, and the survival functions of the fitted NMW and RNMW distributions.

It is clear that both the NMW and RNMW distributions provide adequate fits.
Both have very small K-S values (0.088 and 0.092, respectively).
The NMW distribution has the larger log-likelihood of -212.9.
However, the RNMW distribution has the smaller values for AIC, BIC and CAIC.
The likelihood ratio test statistic for testing $H_{0}:\theta=\gamma=\frac{1}{2}$ versus $H_{1}: H_{0}$ is false is $\omega=1.436$
and the corresponding $p$-value is 0.488, so there is no evidence  to reject $H_{0}$.
Hence, the NMW distribution does not improve significantly on the fit of the RNMW distribution.

The plots of the empirical TTT-transform, TTT-transforms of the fitted NMW and RNMW distributions,
the nonparametric hazard rate function, and the hazard rate functions of the fitted
NMW and RNMW distributions are shown in Figures \ref{a1234}c and \ref{a1234}d.
It is clear that the RNMW distribution provides as good a fit as the NMW distribution.

\begin{table}
\caption{MLEs of parameters, standard errors, AIC, BIC and CAIC for Aarset data.}
\begin{center}
\scalebox{0.8}{\begin{tabular}[h]{  l c c c c c c c c} \hline
    Model &  $\widehat{\alpha}$ & $\widehat{\beta}$ & $\widehat{\gamma}$ & $\widehat{\theta}$ & $\widehat{\lambda}$&AIC&BIC&CAIC \\ \hline
NMW & $0.071 $ & $7.015\times 10^{-8}$ & $0.016$ & $0.595$ & $0.197$& $435.8$ & $445.4$ &$437.2$ \\
& (0.031) & ($1.501\times 10^{-7}$) & (3.602) & (0.128) & (0.184)&&& \\\hline
RNMW & $0.102$ & $ 3.644\times 10^{-8}$ & $\frac{1}{2}$ & $\frac{1}{2}$ & $0.180$& $433.1$ & $439.0$ & $433.8$\\
& (0.019) & ($6.089\times 10^{-8}$) & $-$ & $-$ & (0.020)&&& \\\hline
    \end{tabular}}
\end{center}
\label{MLEs Aarst}
\end{table}

\begin{table}
\caption{K-S statistics for models fitted to Aarset data.}
\begin{center}
 \scalebox{0.8}{\begin{tabular}[h]{l l l l l l l l l l l l l l l l l}
\hline
\textsf{Model} &&&&&&&& K-S  &&&&&&&&  \\ \hline
NMW            &&&&&&&&0.088 &&&&&&&&  \\
RNMW           &&&&&&&& 0.092&&&&&&&&  \\ \hline
\end{tabular}}
\end{center}
\label{KS_AIC_Aarst}
\end{table}

The variance-covariance matrix for the fitted RNMW distribution is
\begin{eqnarray*}
I^{-1}=\left[
\begin{array}{lll}
5.203\times10^{-4} & -3.697\times10^{-9} & 1.214\times10^{-3} \\
-3.697\times10^{-9}& 9.078\times10^{-14} & -2.994\times10^{-8}\\
1.214\times10^{-3}& -2.994\times10^{-8} &9.882\times10^{-3}
\end{array}
\right].
\end{eqnarray*}
So, approximate $95$ percent confidence intervals for the parameters $\alpha $, $\beta$ and $\lambda $
are $\left[ 0.064,0.14\right]$, $\left[ 0,1.558\times 10^{-7}\right]$ and
$\left[0.219\times 10^{-5},0.142\right]$, respectively.

\subsubsection{Kumar data}

\begin{figure}
 \centering
\begin{tabular}{cc}
  \includegraphics[width=7.25cm]{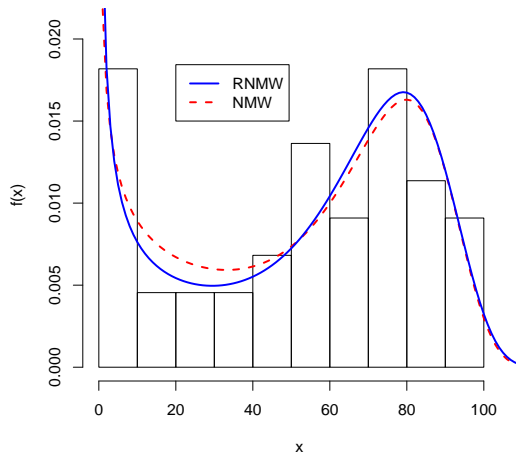} &
  \includegraphics[width=7cm]{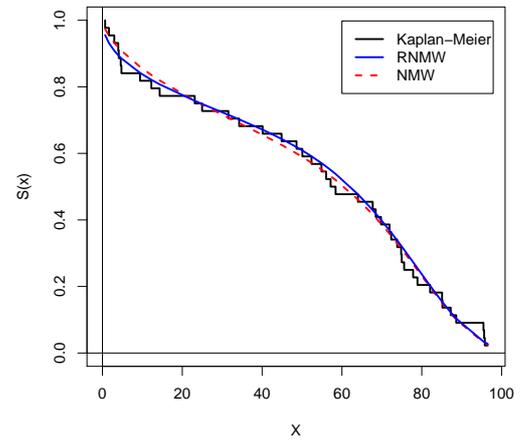}\\
  (a) & (b)\\
  \includegraphics[width=8.5cm]{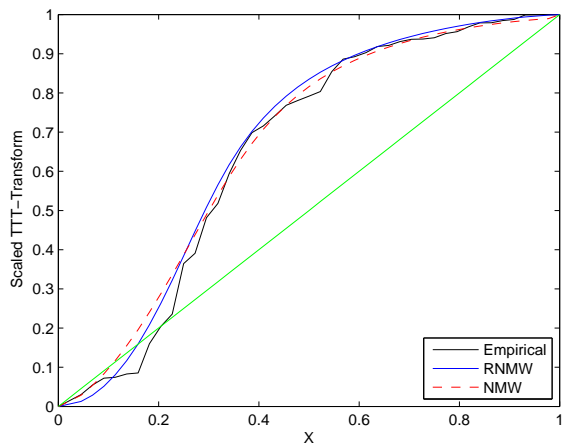} &
  \includegraphics[width=7cm]{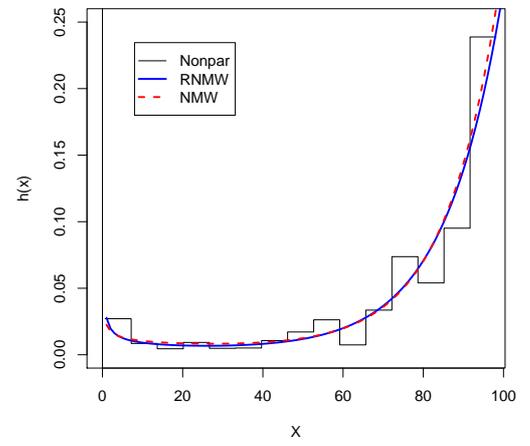}
  \\
  (c) & (d)
  \end{tabular}
\caption{For Kumar data: (a) Histogram and fitted PDFs;
(b) Empirical and fitted survival functions;
(c) Empirical and fitted TTT-transforms;
(c) Nonparametric and fitted hazard rate functions.}
\label{b1234}
\end{figure}

Kumar {\em et al.} \cite{Kumar_data} presented data consisting of
times between failures (TBF) in days of load-haul-dump machines (LHD)
used to pick up rock or waste.
This data set also exhibits a bathtub shaped hazard rate function as shown in Figures \ref{b1234}c and \ref{b1234}d.

Tables \ref{MLEs Kumar} and \ref{KS_AIC_Kumar}
show the MLEs of the parameters, corresponding standard errors, AIC, BIC, CAIC and the K-S test statistics.
Both distributions (NMW and RNMW) provide adequate fits.
The log-likelihood is larger for the NMW distribution.
That for the RNMW distribution is only slightly smaller.
The K-S statistic is larger for the NMW distribution.
However, the RNMW distribution has the smaller AIC, BIC and CAIC values.

\begin{table}
\caption{MLEs of parameters, standard errors, AIC, BIC and CAIC for Kumar data.}
\begin{center}
    \scalebox{0.8}{\begin{tabular}[h]{  l c c c c c c c c} \hline
    Model &  $\widehat{\alpha}$ & $\widehat{\beta}$ & $\widehat{\gamma}$ & $\widehat{\theta}$ & $\widehat{\lambda}$&AIC&BIC&CAIC \\ \hline
NMW & $0.034 $ & $4.588\times 10^{-4}$ & $0.423$ & $0.657$ & $0.071$& $410.0$ & $418.9$ &$411.6$ \\
& (0.022) & ($3.794\times 10^{-3}$) & (2.351) & (0.18) & (0.034)&&& \\\hline
RNMW & $0.055$ & $ 5.852\times 10^{-4}$ & $\frac{1}{2}$ & $\frac{1}{2}$ & $0.065$& $406.9$ & $412.2$ & $407.5$\\
& (0.017) & ($6.523\times 10^{-4}$) & $-$ & $-$ & (0.012)&&& \\\hline
    \end{tabular}}
\end{center}
\label{MLEs Kumar}
\end{table}

\begin{table}
\caption{K-S statistics for models fitted to Kumar data.}
\begin{center}
 \scalebox{0.8}{\begin{tabular}[h]{l l l l l l l l l l l l l l l l l}
\hline
\textsf{Model} &&&&&&&& K-S  &&&&&&&&  \\ \hline
NMW            &&&&&&&&0.068&&&&&&&&  \\
RNMW           &&&&&&&& 0.061&&&&&&&& \\ \hline
\end{tabular}}
\end{center}
\label{KS_AIC_Kumar}
\end{table}

Figures \ref{b1234}a-d show that both distributions fit the data adequately.
However,  the log-likelihood ratio statistic
for testing $H_{0}:\theta=\gamma=\frac{1}{2}$ versus $H_{1}: H_{0}$ is false is
$\omega=0.895$ with the corresponding $p$-value of 0.639.
Hence, again there is no evidence that the NMW distribution provides a better fit
than the RNMW distribution.

The variance-covariance matrix for the fitted RNMW distribution is
\begin{eqnarray*}
I^{-1}=\left[
\begin{array}{lll}
2.839\times10^{-4} & -3.968\times10^{-6} & 6.731\times10^{-5} \\
-3.968\times10^{-6}& 4.254\times10^{-7} & -7.836\times10^{-6}\\
6.731\times10^{-5}& -7.836\times10^{-6}&1.498\times10^{-4}
\end{array}
\right].
\end{eqnarray*}
So, approximate $95$ percent confidence intervals for
the parameters $\alpha $, $\beta$ and $\lambda $
are $\left[ 0.022,0.088\right]$, $\left[ 0,1.864\times 10^{-3}\right]$ and $\left[ 0.041,0.089\right]$, respectively.

\subsection{Censored data}

In this section, we show how the RNMW distribution
can be applied in practice for two real censored data sets, one of which is presented here for the first time.

\subsubsection{Drug data}

This data set was collected from a prison in the Middle East in 2011.
It  represents a sample of eighty two prisoners imprisoned for using or selling drugs.
They were all released as part of a general amnesty for prisoners.
We consider the time from release to reoffending to be the failure time.
Of the eighty two prisoners, sixty six were arrested again for abuse or sale of drugs.
After one hundred and eleven weeks, the others were considered to be censored.

Both the NMW and RNMW distributions were fitted to the data.
Tables \ref{MLEs_drug} and \ref{KS_drug}
show the MLEs of the parameters, corresponding standard errors, AIC, BIC, CAIC and the K-S test statistics.
We see that the RNMW distribution has the smaller AIC, BIC and CAIC values.
The K-S statistic values for both distributions are approximately equal to 0.055.

\begin{figure}
 \centering
\begin{tabular}{cc}
  \includegraphics[width=7.25cm]{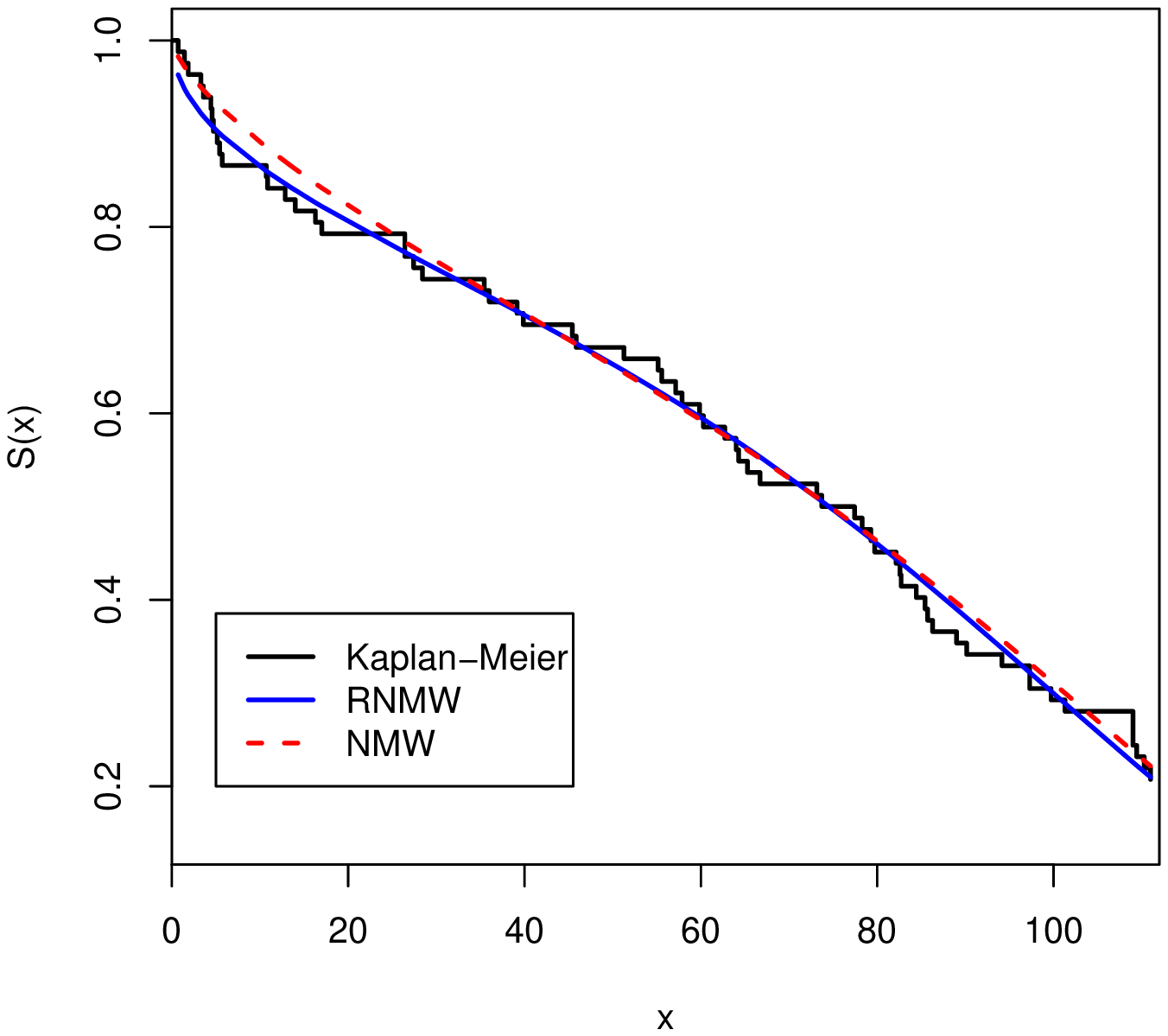} &
  \includegraphics[width=7.25cm]{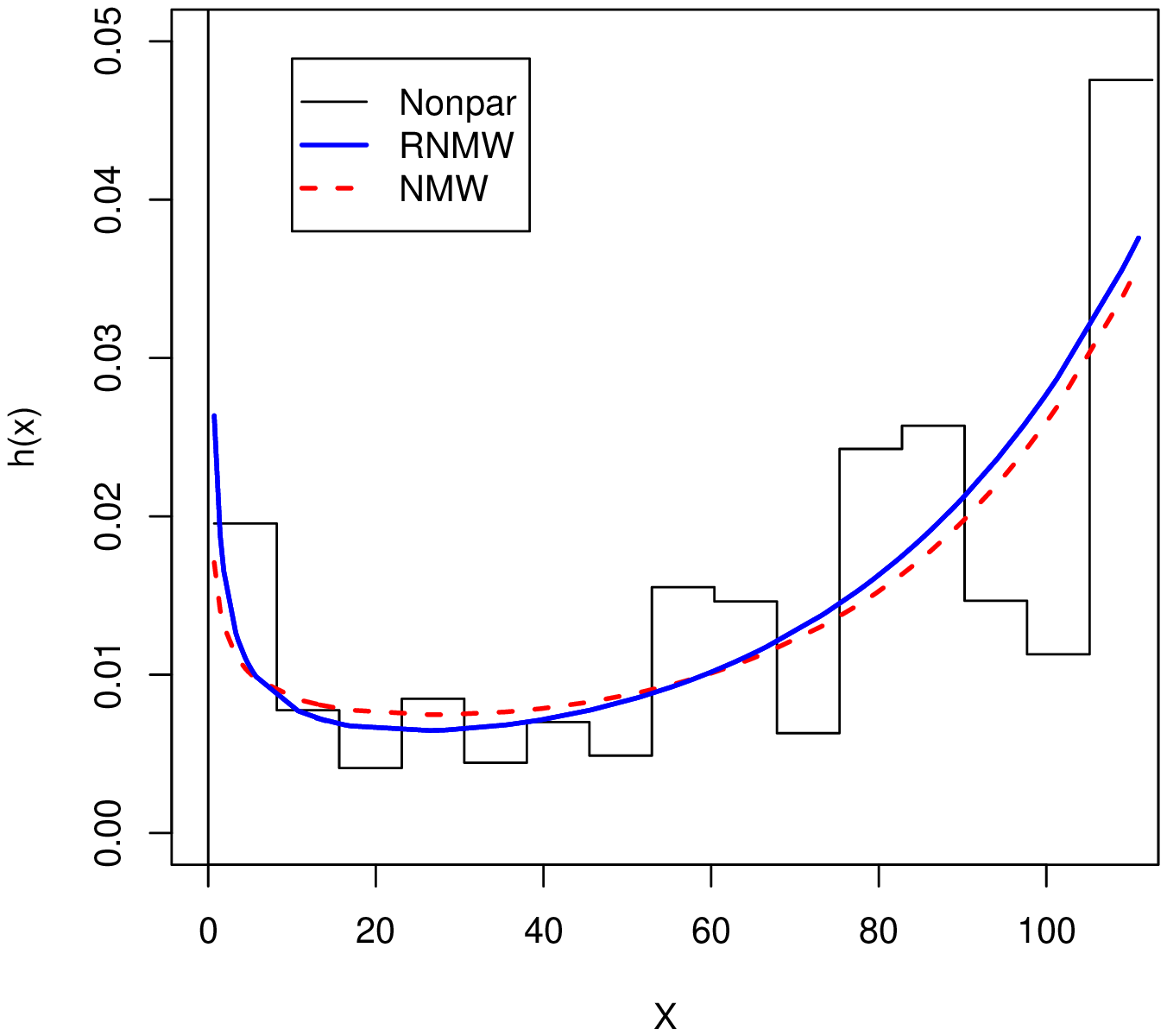}\\
  (a) & (b)\\
  \end{tabular}
\caption {For drug data: (a) Histogram and fitted PDFs;
(b) Empirical and fitted survival functions;
(c) Empirical and fitted TTT-transforms;
(c) Nonparametric and fitted hazard rate functions.}
\label{c1234}
\end{figure}

Figures \ref{c1234}a-d show the histogram of the data, PDFs of the fitted NMW and RNMW distributions,
the empirical survival function, the survival functions of the fitted NMW and RNMW distributions,
the empirical TTT-transform, TTT-transforms of the fitted NMW and RNMW distributions,
the nonparametric hazard rate function, and the hazard rate functions of the fitted NMW and RNMW distributions.
We can see that the RNMW distribution fits the data as well as the NMW distribution.

The log-likelihood ratio statistic
for testing $H_{0}:\theta=\gamma=\frac{1}{2}$ versus $H_{1}: H_{0}$ is false is
$\omega=1.496$ with the corresponding $p$-value of 0.473.
Hence, again there is no evidence that the NMW distribution provides a better fit
than the RNMW distribution.

\begin{table}
\caption{MLEs of parameters, standard errors, AIC, BIC and CAIC for drug data.}
\begin{center}
    \scalebox{0.8}{\begin{tabular}[h]{  l c c c c c c c c} \hline
    Model &  $\widehat{\alpha}$ & $\widehat{\beta}$ & $\widehat{\gamma}$ & $\widehat{\theta}$ & $\widehat{\lambda}$&AIC&BIC&CAIC \\ \hline
NMW &0.019 &$2.954\times10^{-3}$&0.484&0.735&0.031&700.9&713.0&701.7\\
&($9.426\times10^{-3}$)&$(3.656\times10^{-3}$)& (0.504)&(0.159) &(0.019)&&&\\\hline
RNMW &0.038& $ 5.863\times 10^{-3}$ & $\frac{1}{2}$ & $\frac{1}{2}$ & $0.026$&698.4 & 705.6& 698.7\\
& (0.017) & ($6.584\times 10^{-3}$) & $-$ & $-$ & ($9.319\times10^{-3}$)&&& \\\hline
\end{tabular}}
\end{center}
\label{MLEs_drug}
\end{table}

\begin{table}
\caption{K-S statistics for models fitted to the drug data.}
\begin{center}
 \scalebox{0.8}{\begin{tabular}[h]{l l l l l l l l l l l l l l l l l}
\hline
\textsf{Model} &&&&&&&& K-S  &&&&&&&&  \\ \hline
NMW            &&&&&&&&0.0550&&&&&&&&  \\
RNMW           &&&&&&&& 0.0553&&&&&&&&  \\ \hline
\end{tabular}}
\end{center}
\label{KS_drug}
\end{table}

The variance-covariance matrix for the fitted RNMW distribution is
\begin{eqnarray*}
I^{-1}=\left[
\begin{array}{lll}
2.801\times10^{-4} & -8.519\times10^{-5} & 1.138\times10^{-4} \\
-8.519\times10^{-5}& 4.335\times10^{-5} & -6.037\times10^{-5}\\
1.138\times10^{-4}&  -6.037\times10^{-5} &8.684\times10^{-5}
\end{array}
\right].
\end{eqnarray*}
So, approximate $95$ percent confidence intervals for the parameters
$\alpha $, $\beta$ and $\lambda $
are $\left[ 5.553\times 10^{-3},0.071\right]$, $\left[ 0,0.019\right]$ and $\left[8.137\times 10^{-3},0.045\right]$, respectively.

\subsubsection{Serum-reversal data}

Serum-reversal data consists of serum-reversal time in days of one hundred and forty eight
children contaminated
with HIV from vertical transmission at the university hospital of the Ribeirão Preto School of
Medicine (Hospital das Clínicas da Faculdade de Medicina de Ribeirão Preto)
from 1986 to 2001 \cite{Serum_data1}, \cite{Serum_data2}.

Tables \ref{MLEs_serum} and \ref{KS_serum}
show the MLEs of the parameters, corresponding standard errors, AIC, BIC, CAIC and the K-S test statistics.
We see that the RNMW distribution has smaller values for AIC, BIC and CAIC.
The K-S statistic is only slightly larger for the NMW distribution.

\begin{figure}
 \centering
\begin{tabular}{cc}
  \includegraphics[width=7.25cm]{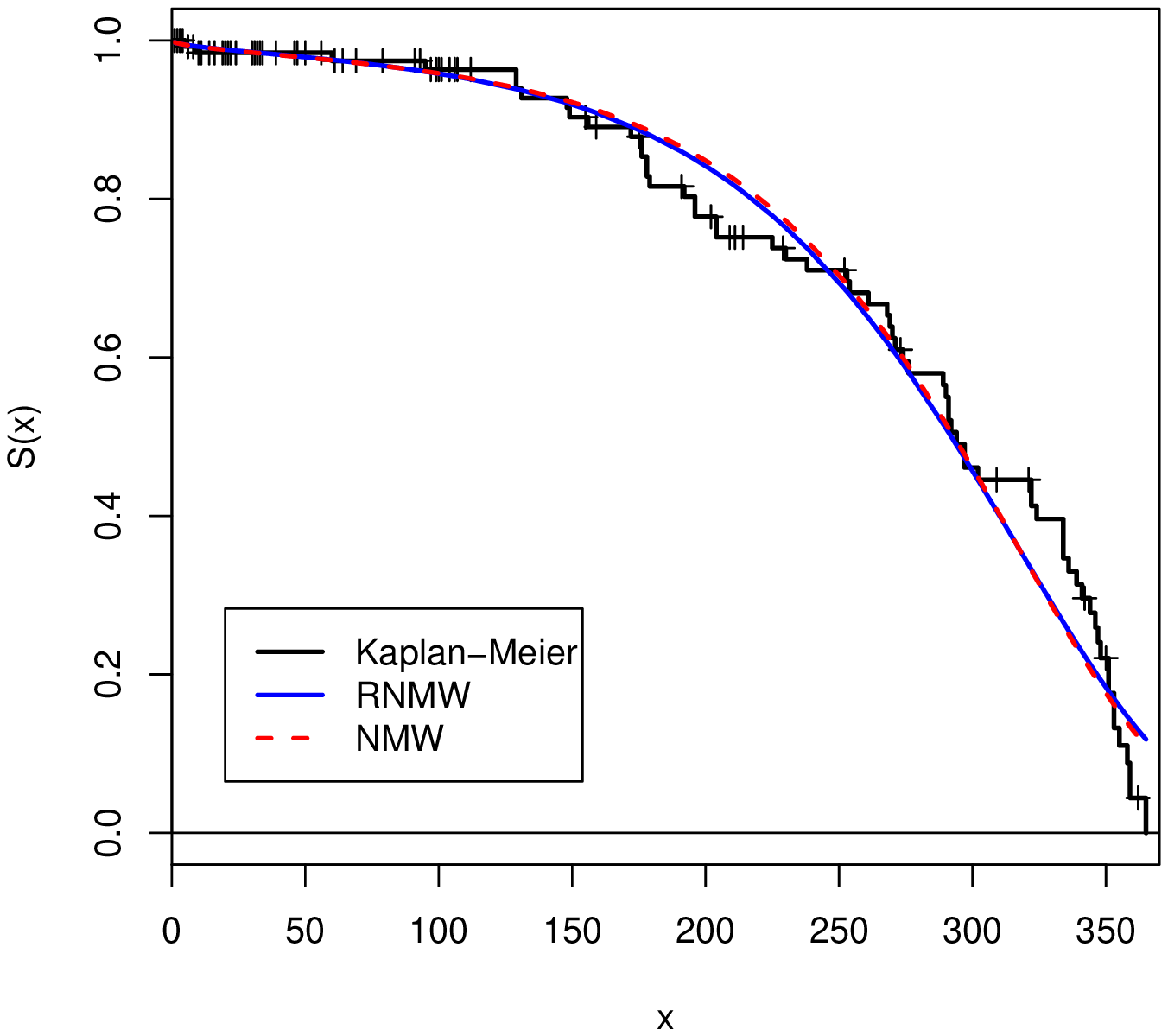} &
  \includegraphics[width=7.25cm]{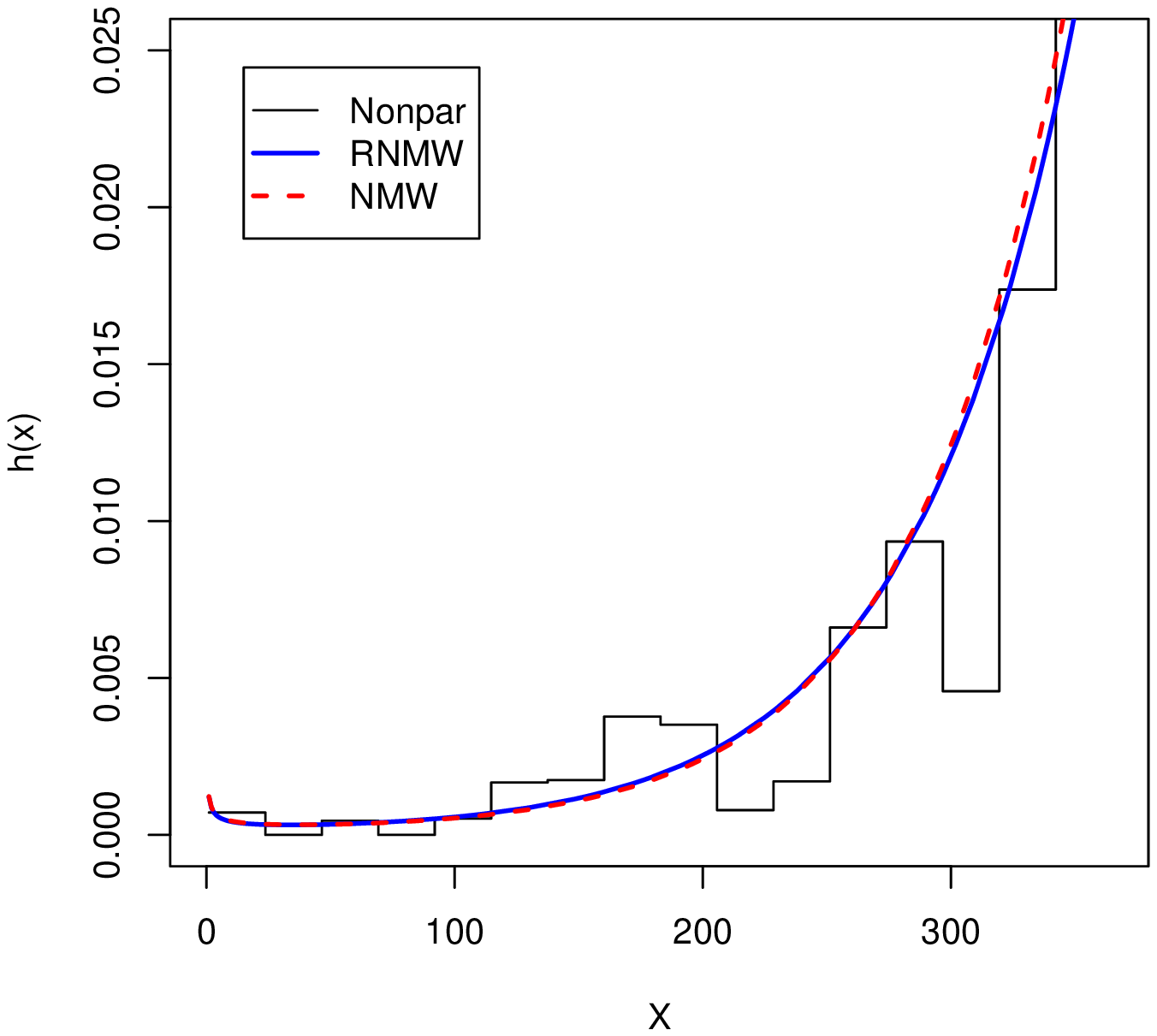}\\
  (a) & (b)\\
  \end{tabular}
\caption {For serum data: (a) Histogram and fitted PDFs;
(b) Empirical and fitted survival functions;
(c) Empirical and fitted TTT-transforms;
(c) Nonparametric and fitted hazard rate functions.}
\label{d1234}
\end{figure}

Figures \ref{d1234}a-d show that both distributions fit the data adequately.
However, the log-likelihood ratio statistic
for testing $H_{0}:\theta=\gamma=\frac{1}{2}$ versus $H_{1}: H_{0}$ is false is
$\omega=0.194$ with the corresponding $p$-value of 0.908.
Hence, again there is no evidence that the NMW distribution provides a better fit
than the RNMW distribution.

\begin{table}
\caption{MLEs of parameters, standard errors, AIC, BIC and CAIC for serum-reversal data.}
\begin{center}
\scalebox{0.8}{\begin{tabular}[h]{  l c c c c c c c c} \hline
Model &  $\widehat{\alpha}$ & $\widehat{\beta}$ & $\widehat{\gamma}$ & $\widehat{\theta}$ &
$\widehat{\lambda}$&AIC&BIC&CAIC \\
\hline
NMW &$1.74\times10^{-3}$ &$6.141\times10^{-4}$&0.542&0.438&0.015&783.8&798.8&803.8 \\
&($3.115\times10^{-3}$)&$(2.274\times10^{-3}$)&(0.507) &(0.763) &($3.764\times10^{-3}$)&&&\\\hline
RNMW &$1.799\times10^{-3}$& $ 5.955\times 10^{-4}$ & $\frac{1}{2}$ & $\frac{1}{2}$ & $0.014$&780.1&789.1&792.1\\
& ($1.971\times10^{-3}$) & ($4.174\times 10^{-4}$) & $-$ & $-$ & ($2.061\times10^{-3}$)&&& \\\hline
    \end{tabular}}
\end{center}
\label{MLEs_serum}
\end{table}

\begin{table}
\caption{K-S statistics for models fitted to serum-reversal data.}
\begin{center}
 \scalebox{0.8}{\begin{tabular}[h]{l l l l l l l l l l l l l l l l l}
\hline
\textsf{Model} &&&&&&&& K-S  &&&&&&&&  \\ \hline
NMW            &&&&&&&& 0.115 &&&&&&&&  \\
RNMW           &&&&&&&& 0.107&&&&&&&&   \\ \hline
\end{tabular}}
\end{center}
\label{KS_serum}
\end{table}

The variance-covariance matrix for the fitted RNMW distribution is
\begin{eqnarray*}
I^{-1}=\left[
\begin{array}{lll}
3.886\times10^{-6} & -4.245\times10^{-7} & 1.990\times10^{-6} \\
-4.245\times10^{-7}& 1.742\times10^{-7} & -8.445\times10^{-7}\\
1.990\times10^{-6}&-8.445\times10^{-7}&4.246\times10^{-6}
\end{array}
\right].
\end{eqnarray*}
So, approximate $95$ percent confidence intervals for the parameters $\alpha $, $\beta$ and $\lambda $
are $\left[ 0,5.662\times 10^{-3}\right]$, $\left[ 0,1.414\times 10^{-3}\right]$ and $\left[0.010,0.080\right]$,
respectively.

\section{Conclusions and further discussion}

The NMW distribution introduced by Almalki and Yuan \cite{meNMW} has been simplified
with its five parameters reduced to three.
The simplified distribution has been referred to as the RNMW distribution.
We have studied several analytical properties of the RNMW distribution
and shown it to be a tractable distribution.
We have also shown that the RNMW distribution provides excellent fits to four real data sets:
two of them are complete data sets and the other two are censored.
By means of the likelihood ratio test, we have shown that the fit of the NMW distribution
is not significantly better than that of the RNMW distribution.
So, the RNMW distribution retains the same flexibility of the NMW distribution
and yet the estimation for the former is much easier.

\begin{figure}[here]
 \centering
  \includegraphics[width=11cm,height=7.5cm]{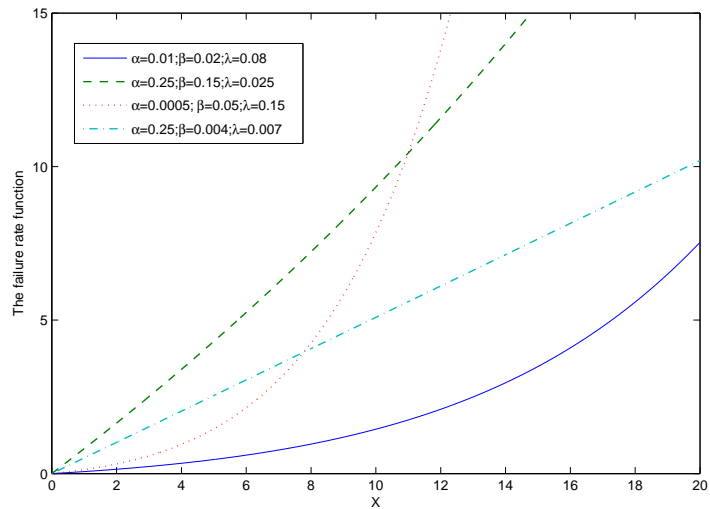}\\
  \caption{Hazard rate functions of the reduced distribution.}
\label{hazrdsRMW_IFR}
\end{figure}

The RNMW distribution has an exclusive bathtub shaped hazard rate function.
Other hazard rates can be obtained from the NMW distribution.
For example, setting $\gamma = \theta = 2$ we obtain
\begin{eqnarray*}
h(x) = 2\alpha x+\beta (2 +\lambda x)x e^{\lambda x}
\end{eqnarray*}
for $x > 0$, which is an increasing function of $x$, see Figure \ref{hazrdsRMW_IFR}.

\appendix
\section{Appendix}

The moment generating function of $X$ is
\begin{eqnarray}
M_{X}(t)
&=&
E\left(e^{tx}\right)
\nonumber
\\
&=&
\int_{0}^{\infty}e^{tx}f(x)dx
\nonumber
\\
&=&
\int_{0}^{\infty}e^{tx}dF(x)
\nonumber
\\
&=&
1+\int_{0}^{\infty}te^{tx}e^{-\alpha \sqrt{x}-\beta \sqrt{x}e^{\lambda x}}dx
\nonumber
\\
&=&
1+\sum_{n,m,k=0}^{\infty}\frac{(-\beta)^{n}(n\lambda)^{m}t^{k+1}}{i!j!k!}I,
\nonumber
\end{eqnarray}
where $I=\int_{0}^{\infty}x^{\frac{n}{2}+m+k}e^{-\alpha \sqrt{x}}dx$.
Using gamma-integral formula, we obtain
\begin{eqnarray*}
M_{X}(t)=1+2\sum_{n,m,k=0}^{\infty}\frac{(-\beta)^{n}(n\lambda)^{m}t^{k+1}}
{i!j!k!}\left[ \frac{\Gamma \left( n+2(m+k)+2\right)}{\alpha^{n+2(m+k)+2}} \right].
\end{eqnarray*}

\section{Appendix}

The elements of the observed information matrix $I(\vartheta)$ for the $RNMW(\alpha, \beta, \lambda)$
for complete data are
\begin{eqnarray*}
&&
\ell_{\alpha \alpha} = -\sum\limits_{i=1}^{d}\left[ \frac{h_{\alpha}
\left( x_{i}; \underline{\vartheta} \right)}{h\left( x_{i};\underline{\vartheta} \right)}\right]^{2},
\\
&&
\ell_{\alpha \beta} = -\sum\limits_{i=1}^{d}
\frac {h_{\alpha} \left( x_{i}; \underline{\vartheta} \right)
h_{\beta} \left( x_{i}; \underline{\vartheta} \right)}{h \left(x_{i};\underline{\vartheta}\right)^{2}},
\\
&&
\ell_{\alpha \lambda} = -\sum\limits_{i=1}^{d}\frac{h_{\alpha} \left( x_{i}; \underline{\vartheta} \right)
h_{\lambda} \left( x_{i};\underline{\vartheta} \right)}{h \left( x_{i};\underline{\vartheta} \right)^{2}},
\\
&&
\ell_{\beta \beta} = -\sum\limits_{i=1}^{d}\left[
\frac {h_{\beta} \left( x_{i}; \underline{\vartheta} \right)}
{h \left( x_{i}; \underline{\vartheta} \right)}\right]^{2},
\\
&&
\ell_{\beta \lambda} = -\sum\limits_{i=1}^{d}
\frac {h \left( x_{i};\underline{\vartheta} \right)
h_{\beta \lambda} \left( x_{i};\underline{\vartheta} \right) -
h_{\beta} \left( x_{i}; \underline{\vartheta} \right)
h_{\lambda} \left( x_{i}; \underline{\phi} \right)}{h \left( x_{i};\underline{\vartheta} \right)^{2}},
\\
&&
\ell_{\lambda \lambda} = \sum\limits_{i=1}^{d}\left[ \frac {h \left(x_{i};\underline{\vartheta}\right)
h_{\lambda \lambda} \left(x_{i};\underline{\vartheta}\right) - h_{\lambda} \left( x_{i};\underline{\vartheta} \right)^{2}}
{h \left( x_{i};\underline{\vartheta} \right)^{2}} - \beta \sqrt{x_{i}^{5}} e^{\lambda x_{i}}\right],
\end{eqnarray*}
where
\begin{eqnarray*}
&&
h_{\alpha} \left( x_{i};\underline{\vartheta} \right) = \frac {1}{2\sqrt{x_{i}}},
\\
&&
h_{\beta} \left( x_{i};\underline{\vartheta} \right) = \frac {\left( 0.5 +\lambda x_{i} \right) e^{\lambda x_{i}}}{\sqrt{x_{i}}},
\\
&&
h_{\lambda} \left( x_{i}; \underline{\vartheta} \right) =
\beta \sqrt{x_{i}}\left(\frac{3}{2} +\lambda x_{i}\right)e^{\lambda x_{i}},
\\
&&
h_{\beta \lambda} \left( x_{i}; \underline{\vartheta} \right) =
\sqrt{x_{i}}\left(\frac{3}{2} +\lambda x_{i}\right)e^{\lambda x_{i}},
\\
&&
h_{\lambda \lambda} \left( x_{i}; \underline{\vartheta} \right) =
\beta \sqrt{x_{i}}\left(\frac{5}{2} +\lambda x_{i}\right)e^{\lambda x_{i}}.
\end{eqnarray*}

\section{Appendix}

The elements of the observed information matrix $I(\vartheta)$ for the $RNMW(\alpha, \beta, \lambda)$
for censored data are
\begin{eqnarray*}
&&
I_{\alpha \alpha} = -\sum\limits_{i=1}^{d}\left[ \frac{h_{\alpha} \left( x_{i};\underline{\vartheta} \right)}
{h \left( x_{i};\underline{\vartheta} \right)}\right]^{2},
\\
&&
I_{\alpha \beta} = -\sum\limits_{i=1}^{d}
\frac {h_{\alpha} \left( x_{i}; \underline{\vartheta} \right)
h_{\beta} \left( x_{i};\underline{\vartheta} \right)}{h\left( x_{i};\underline{\vartheta} \right)^{2}},
\\
&&
I_{\alpha \lambda} = -\sum\limits_{i=1}^{d}
\frac {h_{\alpha} \left( x_{i}; \underline{\vartheta} \right)
h_{\lambda} \left( x_{i};\underline{\vartheta} \right)}
{h\left( x_{i};\underline{\vartheta} \right)^{2}},
\\
&&
I_{\beta \beta} = -\sum\limits_{i=1}^{d}\left[
\frac {h_{\beta} \left( x_{i}; \underline{\vartheta} \right)}
{h \left(  x_{i};\underline{\vartheta} \right)}\right]^{2},
\\
&&
I_{\beta \lambda} = -\sum\limits_{i=1}^{d}
\frac {h \left( x_{i};\underline{\vartheta} \right)
h_{\beta \lambda} \left( x_{i};\underline{\vartheta} \right) -
h_{\beta} \left( x_{i};\underline{\vartheta} \right)
h_{\lambda} \left( x_{i};\underline{\phi} \right)}
{h \left( x_{i};\underline{\vartheta} \right)^{2}} -
\sum\limits_{i\in C}\sqrt{x_{i}^3}e^{\lambda x_{i}},
\\
&&
I_{\lambda \lambda} = \sum\limits_{i=1}^{d}\left[ \frac {h \left( x_{i}; \underline{\vartheta} \right)
h_{\lambda \lambda} \left( x_{i};\underline{\vartheta} \right) -
h_{\lambda} \left( x_{i};\underline{\vartheta} \right)^{2}}
{h \left( x_{i};\underline{\vartheta} \right)^{2}} - \beta\sqrt{x_{i}^5}e^{\lambda x_{i}}\right]
\\
&&
\qquad \qquad
-\beta \sum\limits_{i\in C}\sqrt{x_{i}^5}e^{\lambda x_{i}},
\end{eqnarray*}
where
\begin{eqnarray*}
&&
h_{\alpha} \left( x_{i};\underline{\vartheta} \right) = \frac {1}{2\sqrt{x_{i}}},
\\
&&
h_{\beta} \left( x_{i};\underline{\vartheta} \right) = \frac {\left( 0.5 +\lambda x_{i} \right)e^{\lambda x_{i}}}{\sqrt{x_{i}}},
\\
&&
h_{\lambda} \left( x_{i}; \underline{\vartheta} \right) = \beta \sqrt{x_{i}}\left( \frac{3}{2} +\lambda
x_{i}\right)e^{\lambda x_{i}},
\\
&&
h_{\beta \lambda} \left( x_{i};\underline{\vartheta} \right) =
\sqrt{x_{i}}\left( \frac{3}{2} +\lambda x_{i}\right)e^{\lambda x_{i}},
\\
&&
h_{\lambda \lambda} \left( x_{i};\underline{\vartheta} \right) =
\beta\sqrt{x_{i}^3}\left(\frac{5}{2} +\lambda x_{i}\right)e^{\lambda x_{i}}.
\end{eqnarray*}

\end{document}